\documentstyle[floats,prl,aps,epsf,preprint]{revtex}
\begin{document}
\bibliographystyle{plain}
\title{Emissivities for the various Graviton Modes in the 
Background of the Higher-Dimensional Black Hole} 
\author{ 
D. K. Park\footnote{Email:dkpark@hep.kyungnam.ac.kr 
}}
\address{Department of Physics, Kyungnam University,
Masan, 631-701, Korea.}
\date{\today}
\maketitle

\begin{abstract}
The Hawking emissivities for the scalar-, vector-, and tensor-mode bulk
gravitons are computed in the full range of the graviton's energy by 
adopting the analytic continuation numerically when the spacetime background 
is $(4+n)$-dimensional non-rotating black hole. The total emissivity for the
gravitons is only $5.16\%$ of that for the spin-$0$ field when there is 
no extra dimension. However, this ratio factor increases rapidly when the
extra dimensions exist. For example, 
this factor becomes $147.7\%$, $595.2\%$ and 
$3496\%$ when the number of extra dimensions is $1$, $2$ and 
$6$, respectively. This fact indicates that the Hawking radiation for the 
graviton modes becomes more and more significant and dominant  
with increasing the number of extra dimensions.
\end{abstract}

\newpage
Recent quantum gravity such as string theories\cite{polchin98} or 
brane-world scenario\cite{bwsc1} generally requires the extra dimensions
to reconcile general relativity with quantum physics. Especially the modern
brane-world scenarios predict the emergence of the TeV-scale gravity, which
opens the possibility to make tiny black holes by high-energy scattering
in the future colliders\cite{hec1}.  
In this reason
much attention is paid recently to the effect of the extra dimensions 
in the black hole physics.

The absorption problem and Hawking radiation for the spin $0$, $1/2$ and $1$
particles in the background of the $(4+n)$-dimensional Schwarzschild black 
hole were explored in Ref.\cite{kanti1}. The numerical calculation supports
the fact that the black holes radiate mainly on the brane. In fact this
was pointed out by Emparan, Horowitz and Myers(EHM) in Ref.\cite{emp00} 
by making use of the higher-dimensional black body radiation. This claim was
also supported in the background of the higher-dimensional 
charged black hole\cite{jung05-1}.

More recently, this issue was re-examined when the situation is different.
If, for example, black hole has a rotation, there is an important factor
we should consider carefully called superradiance\cite{superr1}. The 
superradiance in the background of the higher-dimensional black hole was
examined for the bulk fields\cite{frol1} and brane fields\cite{ida02}.
However, numerical calculation has shown that in spite of the consideration
of the superradiance EHM claim still holds due to the incrediably large
difference in the energy amplification for the brane field and 
bulk field\cite{jung05-3}.

There is an another factor we have not considered thoroughly. This is an 
Hawking radiation for the higher-spin particles like graviton. Since the 
graviton is not generally localized on the brane unlike the usual 
standard model particles, the argument of EHM should be carefully re-checked 
in the graviton emission. Generalizing the Regge-Wheeler method\cite{reg57},
the various gravitational perturbations were studied in Ref.\cite{koda03}
in the background of the higher-dimensional Schwarzschild black hole. Using the
radial equations derived in Ref.\cite{koda03}, the low-energy and high-energy
behaviors for the bulk graviton absorption and emission spectra were recently
studied\cite{corn95}. The graviton emission on the brane is also examined using
an axial perturbation\cite{park05-1}. In Ref.\cite{park05-1} it was argued that
the graviton emission can be dominant one in the Hawking radiation when there
are many 
extra dimensions. We would like to explore this issue again in the 
bulk emission. In the following we will compute the absorption and emission 
spectra for the various graviton modes numerically.  
We will show the emission rates are generally enhanced when the number of 
extra dimensions,{\it say} $n$, increases. However, the increasing rate for 
the gravitons is much larger than that for the spin-$0$ field. For example,
the total emissivities for the gravitons is only $5.16\%$ of that for the
spin-$0$ field in four dimensions. However, this ratio becomes 
$595.2\%$ when $n=2$ and $3496\%$ when $n=6$. Thus the Hawking radiation 
for the higher-spin fields becomes dominant when the extra dimensions exist.

We start with $(4+n)$-dimensional Schwarzschild spacetime whose metric is 
\begin{equation}
\label{metric1}
ds^2 = - h dt^2 + h^{-1} dr^2 + r^2 d \Omega_{n+2}^2
\end{equation}
where $h = 1 - (r_H / r)^{n+1}$ and  the angle part $d \Omega_{n+2}^2$ is a 
spherically symmetric line element in a form
\begin{equation}
\label{angle-part}
d\Omega_{n+2}^2 = d\theta_{n+1}^2 + 
\sin^2 \theta_{n+1} \Bigg[ d\theta_{n}^2 + \sin^2 \theta_n \bigg(
\cdots + \sin^2 \theta_2 \left( d\theta_1^2 + \sin^2 \theta_1 d\varphi^2
           \right) \cdots \bigg) \Bigg].
\end{equation}

It is well-known\cite{koda03} that when $n \neq 0$, 
there are three gravitational metric perturbations, 
{\it i.e.} scalar, vector and tensor perturbations. For the vector and tensor
modes the radial equation reduces to the following Schr\"{o}dinger-like 
expression
\begin{equation}
\label{VT1}
\left( \frac{d^2}{d r_{\ast}^2} + \omega^2 \right) \Psi = V_{VT} \Psi
\end{equation}
where $r_{\ast}$ is a ``tortoise'' coordinate defined $d r / d r_{\ast} = h$ 
and the effective potential $V_{VT}$ is 
\begin{eqnarray}
\label{poten1}
V_{VT}&=& \frac{h}{r^2} \left[ \ell (\ell + n + 1) + \frac{n (n + 2)}{4}
                          - \frac{k (n+2)^2}{4} 
                         \left(\frac{r_H}{r} \right)^{n+1} \right]
                                                               \\  \nonumber
& & k = \left\{ \begin{array}{ll}
                 -1        &      \mbox{for tensor mode}   \\
                  3        &      \mbox{for vector mode}
                 \end{array}
                 \right.
\end{eqnarray}     
with $\ell \geq 2$.

The radial equation for the scalar mode also reduces to the 
Schr\"{o}dinger-like expression. However, the effective potential $V_S$
is comparatively complicate as following
\begin{equation}
\label{poten2}
V_S = \frac{h}{r^2}
\frac{q y^3 + p y^2 + w y + z}{4 [2 m + (n+2) (n+3) y]^2}
\end{equation}
where $y = 1 - h = (r_H / r)^{n+1}$ and 
\begin{eqnarray}
\label{bozo1}
m&=& \ell (\ell + n + 1) - (n + 2)
\hspace{2.0cm}   q = (n+2)^4 (n+3)^2
                                      \\   \nonumber
z&=& 16 m^3 + 4 m^2 (n+2) (n + 4)
                                     \\   \nonumber
p&=& (n+2) (n + 3) \left[4 m (2 n^2 + 5 n + 6) + n (n + 2) (n + 3) (n - 2)
                                             \right]
                                      \\    \nonumber
w&=& -12 m (n + 2) \left[m (n - 2) + n (n + 2) (n + 3) \right]
\end{eqnarray}
with $\ell \geq 2$.

In the $4d$ limit the vector mode corresponds to the gravitational axial
perturbation\cite{vish70}, whose effective potential reduces to
\begin{equation}
\label{axial}
V_A(r) = \frac{h}{r^2} \left[(2 \lambda + 2) - 3 
                                 \left(\frac{r_H}{r}\right) \right],
\end{equation}
and the scalar mode corresponds to the gravitational polar
perturbation\cite{zer70} with
\begin{equation}
\label{polar}
V_P(r) = \frac{h}{(2 \lambda r + 3 r_H)^2}
\left[8 \lambda^2 (\lambda + 1) + 12 \lambda^2 \left(\frac{r_H}{r}\right)
      + 18 \lambda \left(\frac{r_H}{r}\right)^2 + 
        9 \left(\frac{r_H}{r}\right)^3 \right]
\end{equation}
where $\lambda = (\ell - 1) (\ell + 2) / 2$. There is no correspondence of 
the tensor mode in four dimension. The most striking result in the $4d$ 
gravitational metric perturbations is the fact that the effective potentials
$V_A$ and $V_P$, which look completely different from each other, are expressed
as a single equation\cite{chand83}
\begin{equation}
\label{4dfinal}
V_{P,A} (r) = \pm \beta \frac{d f}{d r_*} + \beta^2 f^2 + \kappa f
\end{equation}
where $\beta = 3 r_H$, $\kappa = 4 \lambda (\lambda + 1)$ and 
$f = h / r (2 \lambda r + 3 r_H)$. In fact, this relation was found when 
Newman-Penrose formalism\cite{newman62} is applied to the $4d$ gravitational 
perturbations. Making use of this explicit relation, one can show that the 
effective potentials $V_A$ and $V_P$ have the same transmission coefficient.
This fact also indicates that the absorption and emission spectra for
the vector mode graviton and scalar mode graviton by a Schwarzschild black hole
are exactly same in four dimensions. 

Now, we would like to discuss how the absorption and emission spectra are 
computed. We first consider the vector and tensor modes of the bulk graviton.
Defining the dimensionless parameters $x \equiv \omega r$ and 
$x_H \equiv \omega r_H$, one can rewrite Eq.(\ref{VT1}) in the following
\begin{eqnarray}
\label{radial1}
& &4 x^2 \left( x^{n+1} - x_H^{n+1} \right)^2 
\frac{d^2 \Psi}{d x^2} + 4 (n+1) x_H^{n+1} x \left( x^{n+1} - x_H^{n+1} \right)
\frac{d \Psi}{d x}
                             \\   \nonumber
& & + \left[4 x^{2 n + 4} - \left( x^{n+1} - x_H^{n+1} \right)
            \left\{ (2 \ell + n) (2 \ell + n + 2) x^{n+1} - k (n+2)^2
                    x_H^{n+1} \right\} \right] \Psi = 0.
\end{eqnarray}
One can easily show that if $\Psi$ is a solution of Eq.(\ref{radial1}), 
its complex conjugate $\Psi^{\ast}$ is also solution. Eq.(\ref{radial1}) also
guarantees the Wronskian between them is 
\begin{equation}
\label{wron1}
W[\Psi^{\ast}, \Psi]_x \equiv \Psi^{\ast} \frac{d \Psi}{d x} - \Psi 
           \frac{d \Psi^{\ast}}{d x} = 
\frac{{\cal C} x^{n+1}}{x^{n+1} - x_H^{n+1}}
\end{equation}
where ${\cal C}$ is an integration constant.

Now, we would like to consider the solution of Eq.(\ref{radial1}), which is 
convergent in the near-horizon regime, {\it i.e.} $x \sim x_H$. Since 
$x = x_H$ is a regular singular point, this solution can be derived by a 
series expression as following
\begin{equation}
\label{nearh1}
{\cal G}_{n,\ell} (x, x_H) = e^{\lambda_n \ln |x - x_H|}
\sum_{N=0}^{\infty} d_{\ell,N} (x - x_H)^N.
\end{equation}
Inserting Eq.(\ref{nearh1}) into (\ref{radial1}) easily yields
\begin{equation}
\label{lambdafac}
\lambda_n = -i \frac{x_H}{n+1}.
\end{equation}
The sign of $\lambda_n$ is chosen for the near-horizon solution to be 
ingoing into the black hole. The recursion relation for the coefficients
$d_{\ell,N}$ can be obtained too. Since it is ,of course, $n$-dependent and
lengthy, we will not present it here. The Wronskian (\ref{wron1}) enables us 
to derive
\begin{equation}
\label{wron2}
W[{\cal G}_{n,\ell}^{\ast}, {\cal G}_{n,\ell}]_x = 
\frac{-2 i |g_{n,\ell}|^2 x^{n+1}}{x^{n+1} - x_H^{n+1}}
\end{equation}
where $g_{n,\ell} \equiv d_{\ell,0}$.

Next, we would like to discuss the solution of Eq.(\ref{radial1}), which is 
convergent in the asymptotic regime. This is also expressed as a series
form by inverse power of $x$ as following:
\begin{equation}
\label{asymp1}
{\cal F}_{n,\ell (\pm)} (x , x_H) = (\pm i)^{\ell + 1 + \frac{n}{2}} x 
e^{\mp i x} (x - x_H)^{\pm \lambda_n} 
\sum_{N=0}^{\infty} \tau_{N (\pm)} x^{-(N+1)}
\end{equation}
with $\tau_{0 (\pm)} = 1$. The solutions ${\cal F}_{n,\ell (+)}$ and 
${\cal F}_{n,\ell (-)}$ are respectively the ingoing and outgoing solutions.
Using Eq.(\ref{wron1}), one can show easily
\begin{equation}
\label{wron3}
W[{\cal F}_{n,\ell (+)}, {\cal F}_{n,\ell (-)}]_x = 
\frac{2 i x^{n+1}}{x^{n+1} - x_H^{n+1}}.
\end{equation}
The recursion relation for the coefficients $\tau_{N (\pm)}$ is 
straightforwardly derived, which is too lengthy to present here.

Now, we would like to discuss how the coefficient 
$g_{n,\ell}(\equiv d_{\ell,0})$ is related to the scattering amplitude. Since 
the real scattering solution, {\it say} ${\cal R}_{n,\ell}$ should be 
ingoing wave in the near-horizon regime and the mixture of ingoing and 
outgoing waves at the asymptotic regime, one may express it in the form:
\begin{eqnarray}
\label{real-sol}
& &{\cal R}_{n,\ell}\stackrel{x \rightarrow x_H}{\sim} 
g_{n,\ell}(x - x_H)^{\lambda_n}
\left[1 + O(x -x_H)\right]   \\ \nonumber
& &{\cal R}_{n,\ell}\stackrel{x \rightarrow \infty}{\sim} 
\frac{i^{\ell + 1 + \frac{n}{2}} 2^{\frac{n}{2} - 1}}
     { \sqrt{\pi}}
\Gamma \left( \frac{1 + n}{2} \right)
Q_{n,\ell}
                                                \\  \nonumber
& & \hspace{2.0cm} \times
\left[e^{-i x + \lambda_n \ln |x - x_H|} - (-1)^{\ell + \frac{n}{2}}
      S_{n,\ell}(x_H) e^{i x - \lambda_n \ln |x - x_H|} \right]
+ O\left(\frac{1}{x}\right)
\end{eqnarray}
where $S_{n,\ell}(x_H)$ is a scattering amplitude and $Q_{n,\ell}$ is a 
quantity related to the multiplicities ${\cal N}_{n,\ell}$ as following:
\begin{equation}
\label{add1}
Q_{n,\ell} = \frac{2}{(n+4) (n+1)} {\cal N}_{n,\ell}.
\end{equation} 
The multiplicities ${\cal N}_{n,\ell}^{(S)}$, ${\cal N}_{n,\ell}^{(V)}$ and
${\cal N}_{n,\ell}^{(T)}$ for the scalar-, vector- and tensor-mode 
bulk gravitons are given by\cite{corn95} 
\begin{eqnarray}
\label{add2}
{\cal N}_{n,\ell}^{(S)}&=& \frac{(2 \ell + n + 1) (\ell + n) !}
                                {\ell ! (n+1) !}
                                                      \\  \nonumber
{\cal N}_{n,\ell}^{(V)}&=& \frac{\ell (\ell + n + 1) (2 \ell + n + 1) 
                                     (\ell + n - 1)!}
                                {(\ell + 1) ! n!}
                                                      \\   \nonumber
{\cal N}_{n,\ell}^{(T)}&=& \frac{n (n+3) (\ell + n + 2) (\ell - 1)
                                 (2 \ell + n + 1) (\ell + n - 1)!}
                                {2 (\ell + 1)! (n + 1)!}.
\end{eqnarray}

Introducing a phase shift
$\delta_{n,\ell}$ as $S_{n,\ell} \equiv e^{2 i \delta_{n,\ell}}$, one can 
rewrite the second equation of Eq.(\ref{real-sol}) in the form
\begin{eqnarray}
\label{real-sol2}
& &{\cal R}_{n,\ell}\stackrel{x \rightarrow \infty}{\sim} 
\frac{2^{\frac{n}{2}}}{\sqrt{\pi}} 
\Gamma \left(\frac{1 + n}{2} \right)
Q_{n,\ell}
e^{i \delta_{n,\ell}}
                                               \\  \nonumber
& & \hspace{2.0cm} \times
\sin \left[x + i \lambda_n \ln |x - x_H| - \frac{\pi}{2}
           \left(\ell + \frac{n}{2}\right) + \delta_{n,\ell} \right]
+ O\left(\frac{1}{x}\right).
\end{eqnarray}
  
The near-horizon behavior of the real scattering solution, {\it i.e.} the 
first one of Eq.(\ref{real-sol}), implies that the Wronskian
$W[{\cal R}_{n,\ell}^{\ast}, {\cal R}_{n,\ell}]_x$ is same with 
Eq.(\ref{wron2}). However, Eq.(\ref{real-sol2}) implies
\begin{equation}
\label{wron4}
[{\cal R}_{n,\ell}^{\ast}, {\cal R}_{n,\ell}]_x = -i \frac{2^n}{\pi}
\Gamma^2 \left(\frac{1 + n}{2} \right)
Q_{n,\ell}^2
e^{-2 \beta_{n,\ell}} \sinh 2 \beta_{n,\ell}
\frac{x^{n+1}}{x^{n+1} - x_H^{n+1}}
\end{equation}
where $\delta_{n,\ell} = \eta_{n,\ell} + i \beta_{n,\ell}$. Equalizing these
two Wronskians yields a relation between $g_{n,\ell}$ and $\beta_{n,\ell}$
as following:
\begin{equation}
\label{relat1}
|g_{n,\ell}|^2 = \frac{2^{n-2}}{\pi}
\Gamma^2 \left(\frac{1 + n}{2} \right)
Q_{n,\ell}^2
\left(1 - e^{-4 \beta_{n,\ell}} \right).
\end{equation}
Thus one can calculate the transimission coefficient 
$1 - |S_{n,\ell}|^2 = 1 - e^{-4 \beta_{n,\ell}}$ if $g_{n,\ell}$ is known.

Now, we would like to explain how to compute $g_{n,\ell}$. For the explanation 
it is convenient to introduce a new radial solution 
$\tilde{{\cal R}}_{n,\ell}(x)$, which differs from 
${\cal R}_{n,\ell}(x)$ in its normalization in such a way that
\begin{equation}
\label{newra}
\tilde{{\cal R}}_{n,\ell}(x, x_H) \stackrel{x \rightarrow x_H}{\sim}
(x - x_H)^{\lambda_n} \left[1 + O(x -x_H)\right].
\end{equation}
Since ${\cal F}_{n,\ell(\pm)}$ are two linearly independent solutions of 
Eq.(\ref{radial1}), one can express $\tilde{{\cal R}}_{n,\ell}$ as a linear
combination of them as following
\begin{equation}
\label{newra2}
\tilde{{\cal R}}_{n,\ell}(x, x_H) = f_{n,\ell}^{(-)} (x_H) 
{\cal F}_{n,\ell(+)} (x , x_H) + f_{n,\ell}^{(+)} (x_H)
{\cal F}_{n,\ell(-)} (x , x_H)
\end{equation}
where $f_{n,\ell}^{(\pm)}$ are called jost functions. Using Eq.(\ref{wron3}) 
one can compute the jost functions as following
\begin{equation}
\label{jost1}
f_{n,\ell}^{(\pm)} (x_H) = \pm \frac{x^{n+1} - x_H^{n+1}}{2 i x^{n+1}} 
W[{\cal F}_{n,\ell(\pm)}, \tilde{{\cal R}}_{n,\ell}]_x.
\end{equation}
Inserting the explicit expressions of ${\cal F}_{n,\ell(\pm)}$ in 
Eq.(\ref{asymp1}) into Eq.(\ref{newra2}) and comparing it with the second
equation of Eq.(\ref{real-sol}), one can derive the following two 
relations
\begin{eqnarray}
\label{relat2}
S_{n,\ell}(x_H)&=& \frac{f_{n,\ell}^{(+)} (x_H)}
                        {f_{n,\ell}^{(-)} (x_H)}
                                                   \\   \nonumber
f_{n,\ell}^{(-)} (x_H)&=& \frac{2^{\frac{n}{2} - 1}}
                               {\sqrt{\pi} g_{n,\ell}(x_H)}
\Gamma \left(\frac{1 + n}{2}\right)
Q_{n,\ell}.
\end{eqnarray}
Combining Eq.(\ref{relat1}) and (\ref{relat2}), one can express the greybody
factor(or transmission coefficient) of the black hole in terms of the 
jost function in the following
\begin{equation}
\label{jost2}
1 - |S_{n,\ell}|^2 = \frac{1}{|f_{n,\ell}^{(-)}|^2}.
\end{equation}

The partial absorption cross sections for the vector and tensor 
graviton modes by
the $(4+n)$-dimensional Schwarzschild black hole are given by
\begin{equation}
\label{abs1}
\sigma_{n,\ell} = 2^{n+1} \pi^{(n+1) / 2}
\Gamma \left( \frac{3 + n}{2} \right)
Q_{n,\ell}
\frac{r_H^{n+2}}{x_H^{n+2} |f_{n,\ell}^{(-)}|^2}.
\end{equation}
Applying the Hawking formula\cite{hawking75}, one can compute the bulk 
emission rate, {\it i.e.} the energy emitted to the bulk per unit time and
unit energy interval, as following
\begin{equation}
\label{emission1}
\frac{\Gamma}{d \omega} = 
\left[2^{n + 3} \pi^{(n+3) / 2} \Gamma \left( \frac{3 + n}{2} \right)
                    \right]^{-1} (n + 4) (n + 1)
\frac{\omega^{n+3} \sigma_{abs} (\omega)}
     {e^{\omega / T_H} - 1}
\end{equation}
where $\sigma_{abs}$ is a total absorption cross section and $T_H$ is an 
Hawking temperature given by $T_H = (n+1) / 4 \pi r_H$. Thus one can compute
the absorption and emission spectra in the full range of $\omega$ if the 
jost functions are computed.

One can compute the jost functions by matching the near-horizon and 
asymptotic solutions. This is achieved by the analytic 
continuation\cite{jung05-1,jung05-3}
using a solution of Eq.(\ref{radial1}) which is convergent at the 
arbitrary point $x = b$. This solution is also expressed as a series
form as following
\begin{equation}
\label{interme}
\varphi_{n,\ell} (x, x_H) = (x - x_H)^{\lambda_n}
\sum_{N=0}^{\infty} D_N (x - b)^N.
\end{equation}
The recursion for the coefficients $D_N$ can be obtained by inserting 
Eq.(\ref{interme}) into (\ref{radial1}). Since it is too lengthy, we will not
present it here. Using a solution (\ref{interme}) one can increase the 
convergent region for the asymptotic solution from the near-horizon regime 
and decrease for the asymptotic solution from the 
asymptotic regime. Repeating the procedure eventually makes the two 
solutions which have common convergent region. Then one can compute the jost
functions by making use of these two solutions and Eq.(\ref{jost1}).

Now, we would like to discuss the case of the scalar mode graviton. Introducing
the dimensionless parameters $x = \omega r$ and $x_H = \omega r_H$ again, one
can rewrite the radial equation for the scalar mode as
\begin{eqnarray}
\label{scalar-radial}
& &4 x^2 \left(x^{n+1} - x_H^{n+1}\right)^2
\left[2 m x^{n+1} + (n + 2) (n + 3) x_H^{n+1} \right]^2
\frac{d^2 \Psi}{d x^2}
                              \\         \nonumber
& &+ 4 (n + 1) x_H^{n + 1} x \left(x^{n+1} - x_H^{n+1}\right)
\left[2 m x^{n+1} + (n + 2) (n + 3) x_H^{n+1} \right]^2
\frac{d \Psi}{d x}
                              \\        \nonumber
& & \hspace{1.0cm}
\Bigg[ 4 x^{2 n + 4} 
          \left[2 m x^{n+1} + (n + 2) (n + 3) x_H^{n+1} \right]^2 
                               \\        \nonumber
& & \hspace{2.0cm}
         - \left(x^{n+1} - x_H^{n+1}\right)
        \left\{z x^{3 n + 3} + w x_H^{n+1} x^{2 n + 2} + p x_H^{2 n + 2}
               x^{n + 1} + q x_H^{3 n + 3}  \right\}
                                             \Bigg] \Psi = 0.
\end{eqnarray} 
The calculational procedure is exactly same with the case of vector or tensor
mode. All Wronskians in Eq.(\ref{wron1}), (\ref{wron2}) and (\ref{wron3}) as 
well as the multiplication factor in Eq.(\ref{lambdafac}) are exactly
same with the previous case. The differences arise only in the recursion
relations for the coefficients $d_{\ell,N}$, $\tau_{N(\pm)}$ and $D_N$ and 
the multiplicity defined in Eq.(\ref{add2}). 
Although the recursion relations in the scalar mode case are much more 
complicate and lengthy, they do not make any difficulty in the numerical
calculation. In actual numerical computation we summed up the partial modes
for $2 \leq \ell \leq 12$ to plot the absorption and emission spectra of 
the graviton modes.

\begin{figure}[ht!]
\begin{center}
\epsfysize=6.0cm \epsfbox{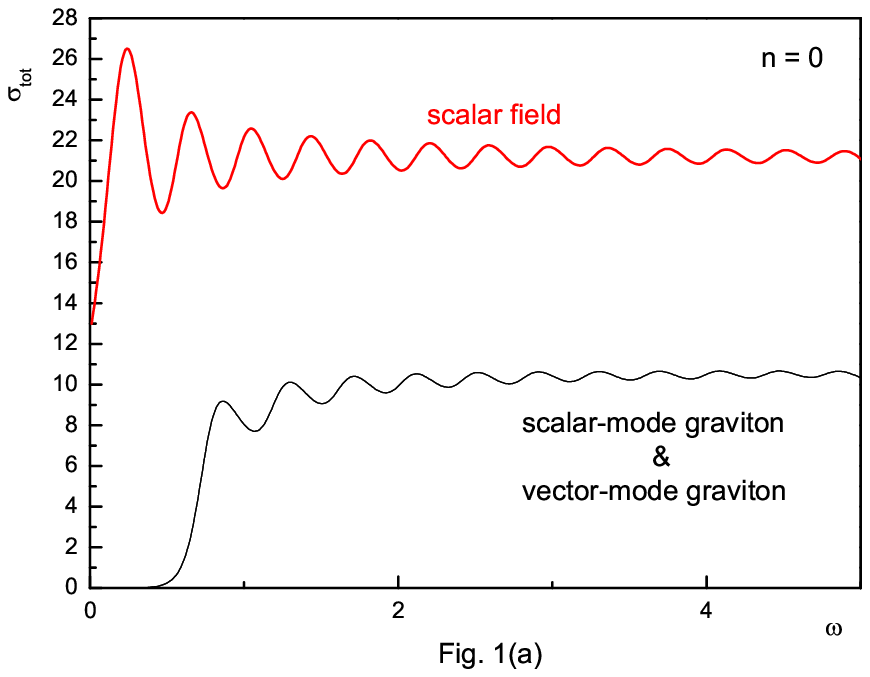}
\epsfysize=6.0cm \epsfbox{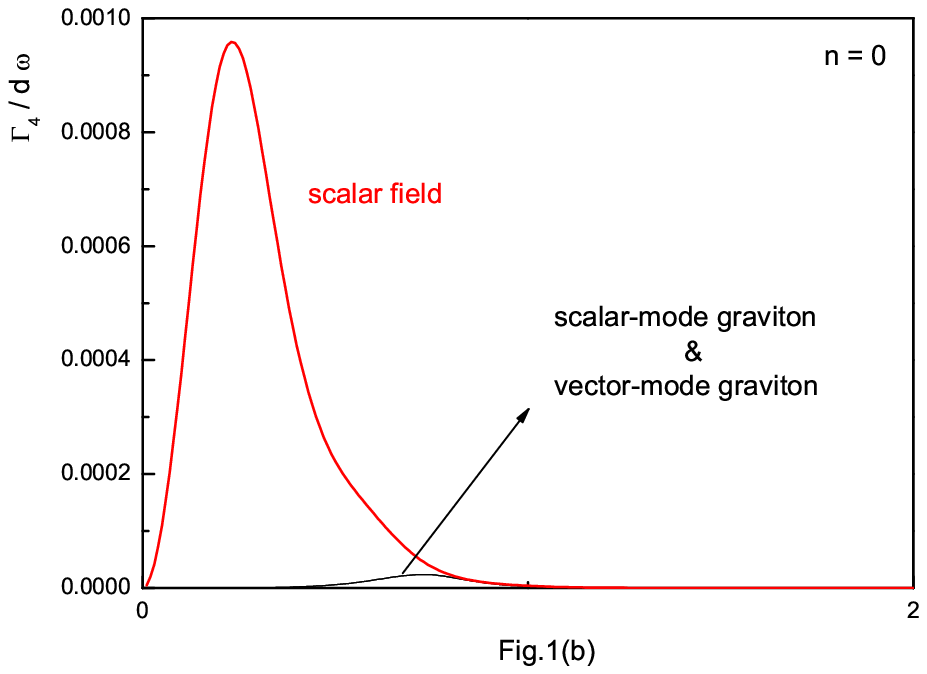}
\caption[fig1]{The absorption (Fig.1(a)) and emission (Fig.1(b)) spectra for 
the scalar- and vector-mode gravitons when $n=0$. 
For a comparision the spectra for the spin-$0$ field are plotted together.
As Eq.(\ref{4dfinal})
implies, the spectra for both gravitons are exactly identical. Fig.1(b)
indicates that the emission rates for the gravitons are negligible compared 
to that for the spin-$0$ field when there is no extra dimension.} 
\end{center}
\end{figure}

Fig.1 shows the total absorption and emission rates for the 
vector(or axial)-mode graviton and scalar(or polar)-mode graviton when
$n=0$. To compare the graviton spectra with those for the low-spin field,
we plot the absorption and emission rates for the spin-$0$ scalar field
together. As Eq.(\ref{4dfinal}) implies, the absorption and emission spectra
for both graviton modes are exactly identical. As Fig.1(b) indicates, the 
emission rates for the gravitons are much smaller than that for the 
spin-$0$ field. The total emission rates for both gravitons are only 
$5.16\%$ of that for the spin-$0$ field.
Thus, in $4d$ the Hawking radiation for the low-spin fields
such as standard model fields is dominant one. However, the situation 
is drastically changed when the extra dimensions exist.

\begin{figure}[ht!]
\begin{center}
\epsfysize=5.5cm \epsfbox{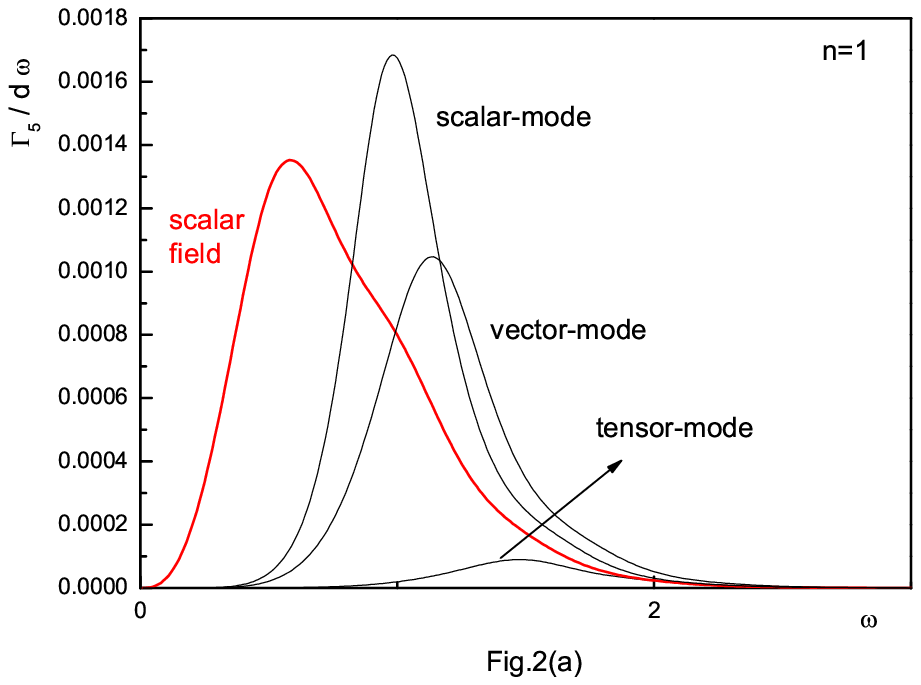}
\epsfysize=5.5cm \epsfbox{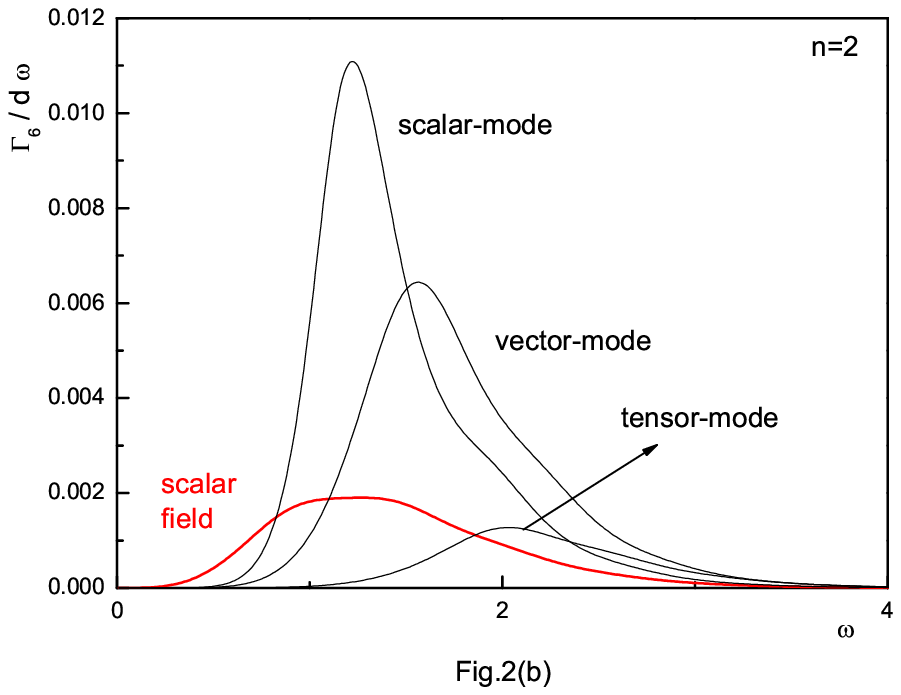}
\epsfysize=5.5cm \epsfbox{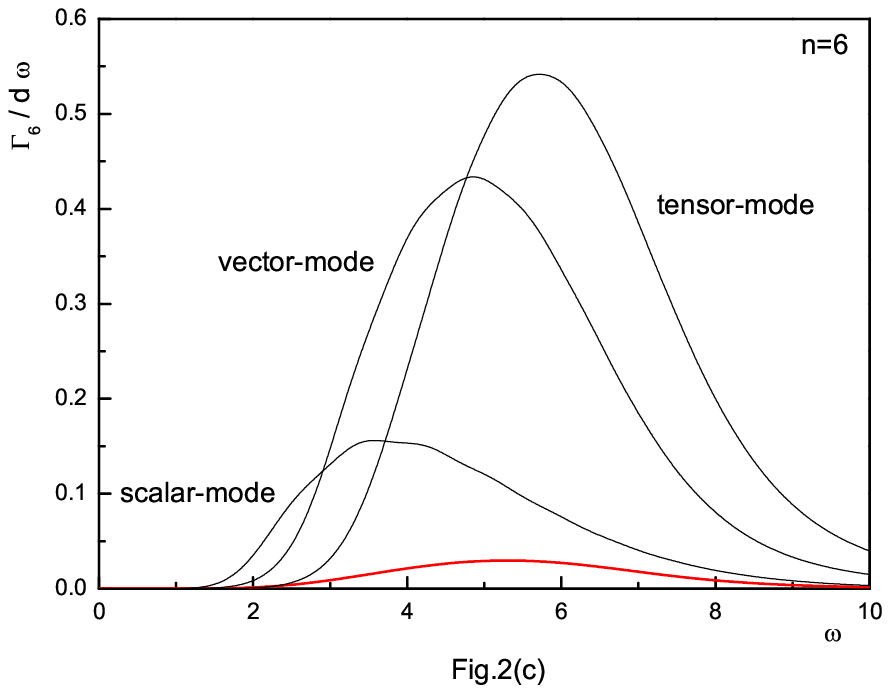}
\caption[fig2]{The emission spectra for the scalar-, vector- and tensor-mode
gravitons when $n=1$ (Fig.2(a)), $n=2$ (Fig.2(b)) and $n=6$ (Fig.2(c)). The
spectra for the spin-$0$ field are plotted together for a comparision.
In general the emissivities for all fields increase with increasing $n$. 
However, the increasing rate for the gravitons are much larger than
that for the spin-$0$ field. Thus, Hawking radiation for the gravitons
becomes more and more significant when the extra dimensions exist.}
\end{center}
\end{figure}

Fig.2 shows that the total emission rates for the scalar-, vector-, and 
tensor-mode gravitons when $n=1$, $2$, and $6$. Like Fig.1 the emission 
spectrum of the spin-$0$ field is plotted together for a comparision. As Fig.2
indicates, the emissivities for all fields generally increase 
with increasing $n$. However,
the increasing rates for the gravitons in the Hawking emissivities with 
increasing $n$ is much larger than that for the spin-$0$ field. 

\vspace{1.0cm}
\begin{center}

\begin{tabular}{l|l|l|l|r} \hline
$ $           & spin-$0$&   scalar-mode / spin-$0$&   vector-mode / spin-$0$ &
                tensor-mode / spin-$0$  \\  \hline \hline
$4d$ &   $0.000297531$   &\hspace{1.0cm}  $0.0258$  &\hspace{1.0cm}  $0.0258$ &
                                                                          \\
$5d$&   $0.00104968$     &\hspace{1.0cm}  $0.8115$  &\hspace{1.0cm} $0.60405$ &
                                        $0.06161$ \hspace{1.0cm}    \\ 
$6d$  &  $0.002679$   &\hspace{1.0cm}  $3.02476$   &\hspace{1.0cm} $2.37418$  
                                        & $0.55351$ \hspace{1.0cm} \\   
$10d$ &  $0.118857$   &\hspace{1.0cm}  $4.94863$   &\hspace{1.0cm}  $13.4278$  
                                        & $16.5793$  \hspace{1.0cm}
                                                                           \\ 
\hline
\end{tabular}

\vspace{0.1cm}
\large{Table I} 
\end{center}
\vspace{1.0cm}

Table 1 shows the relative emissivities when $n=0$, $1$, $2$ and $6$. When
$n = 2$, the emission rate for the scalar mode graviton is almost three
times than that for the spin-$0$ field. When $n = 6$, the emission rate
for the tensor-mode graviton is seventeen times than that for the spin-$0$
field!!! The remarkable fact is that the ratio of the tensor-mode graviton to
the spin-$0$ field increases rapidly with increasing $n$
compared to other graviton modes. 

In this letter we computed the absorption and emission spectra for the 
various modes of the bulk gravitons in the higher-dimensional non-rotating 
black hole background. The total emissivities for the gravitons are only
$5.16\%$ compared to that for the spin-$0$ field in four dimensions. However, 
this ratio increases rapidly when the extra dimensions exist. When, for
example, $n=1$, $2$ and $6$, this ratio goes to $147.7\%$, $595.2\%$ and 
$3496\%$, respectively. This fact indicates that the emission of the 
higher-spin field like gravitons becomes more and more dominant and has a
experimental significance when there are many extra dimensions. Same conclusion
was derived in the emission on the brane\cite{park05-1}. Thus, it is important
in our opinion to re-examine the EHM argument ``{\it black holes radiate
mainly on the brane}'' with considering the Hawking radiation for the 
gravitons. We hope to study this issue in the future.

\vspace{1cm}

{\bf Acknowledgement}:  
This work was supported by the Kyungnam University
Research Fund, 2006.

\end{document}